\documentclass[useAMS,usenatbib]{mn2e}
\usepackage{epsfig}
\usepackage{threeparttable}
\usepackage{amsmath}
\title[Evolution of neutron star + He star binaries]{Evolution of neutron star + He star binaries: an alternative evolutionary channel to intermediate-mass binary pulsars}
\author[W. C. Chen, and W. M. Liu]{ Wen-Cong Chen\thanks{E-mail:
chenwc@pku.edu.cn}, and Wei-Min Liu,\\ School of Physics, Shangqiu
Normal University, Shangqiu 476000, China}
\begin{document}

\date{}

\pagerange{\pageref{firstpage}--\pageref{lastpage}} \pubyear{2011}

\maketitle

\label{firstpage}

\begin{abstract}
It is difficult for intermediate-mass X-ray binaries to form
compact intermediate-mass binary pulsars (IMBPs) with a short
orbital-period ($\la 3~\rm d$), which have a heavy ($\ga 0.4~
M_{\odot}$) CO or ONeMg white dwarf companions. Since neutron star
+ He star binaries may experience common-envelope evolution, they
have some advantage to account for the formation of short
orbital-period IMBPs. In this work, we explore the probability of
IMBPs formed by this evolutionary channel. Using Eggleton's
stellar evolution code, considering that the dead pulsars were
spun up by the accreting material and angular momentum from the He
star companions, we have calculated the evolution of a large
number of neutron star + He star binaries. Our simulated results
indicate that, the NS + He star evolutionary channel can produce
IMBPs with a WD of $\sim0.5 - 1.1~ M_{\odot}$ and an orbital
period of $0.03 - 20$ d, in which pulsars have a spin-period of
$1.4 - 200$ ms. Comparing the calculated results with the
observational parameters (spin period and orbital period) of 9
compact IMBPs, the NS + He star evolutionary channel can account
for the formation of 4 sources. Therefore, NS + He star binaries
offer an alternative evolutionary channel to compact IMBPs.
\end{abstract}


\section{Introduction}
Millisecond pulsars (MSPs) are characterized by weak surface
magnetic fields ($B\sim 10^{8}$ G), short spin periods ($1< P_{\rm
s} < 30$ ms), and nearly circular orbits with low-mass
($0.15~M_{\odot}\la M_{\rm wd}\la 0.4~M_{\odot}$) He white dwarf
(WD) companions. These systems are generally called low-mass
binary pulsars \citep[LMBPs; e.g.][]{taur06}. In contrast to
LMBPs, observations found an intermediate-mass binary pulsar
(IMBP) population \citep{cami96,cami01,edwa01a}, which consists of
a pulsar with a spin period of $\sim 10 - 200$ ms and a heavy ($
M_{\rm wd}\ga 0.4~M_{\odot}$) CO or ONeMg WD.

At present, the formation mechanism of LMBPs is well understood.
In the standard recycling scenario, LMBPs are the evolutionary
products of low-mass X-ray binaries (LMXBs, with hydrogen-rich
donor stars of $\la 1.5~M_{\odot}$) via Case A or B Roche-lobe
overflow (RLOF) \citep[for a review, see][]{bhat91,verb93}. A
pulsar that passed through the so-called death line is spun up to
millisecond period by the accretion of material and angular
momentum when the donor star fills its Roche-lobe
\citep{alpa82,radh82}. Meanwhile, mass accretion onto a neutron
star (NS) leads to magnetic field decay \citep{taam86,roma90}, and
the tidal interaction between the NS and the convective envelope
of the donor star induces a circular orbit \citep{phin92,phin94}.
The orbital-periods of LMBPs usually range from 0.2 days to
several hundreds days, and are strongly correlated with the WD
masses \citep{taur99}.

Comparing with the LMBPs, IMBPs have relatively heavy WDs. In
addition, the spin periods, the inferred surface magnetic fields,
and the orbital eccentricities of the IMBPs are considerably
higher than those of the LMBPs \citep{li02}. These properties
imply that the progenitors and the evolutionary history of IMBPs
are distinct from those of LMBPs. It is generally thought that a
large fraction of IMBPs are descendants of intermediate-mass X-ray
binaries (IMXBs), in which the hydrogen-rich donor stars have a
mass in the interval $\sim 1.5 - 10~M_{\odot}$ \citep{heuv75}. For
example, IMBPs with an orbital-period of $3 - 50$ d evolved from
IMXBs with a donor star of $\sim 2.5 - 5.0~M_{\odot}$ and an
orbital-period of $3 - 10$ d via Case B RLOF
\citep{taur00,pods02,taur12}. In the IMXB evolutionary stage, the
mass transfer rate is much larger than the Eddington accretion
rate of the NS. After the rapid mass transfer, the donor star
evolves into a low-mass star, and the binary begins LMXB
evolutionary phase. The subsequent mass transfer rates are very
low, and trigger accretion disk instability. The accretion process
shows short-lived outbursts separated by long-term quiescence, and
the accretion efficiency markedly declines \citep{li02}. Although
the NS in IMXBs only accretes a transferred amount of material of
$\sim 0.01~M_{\odot}$, it can still be spun up to a short
spin-period of $\sim 10 \rm ms$ \citep{taur12}.

The IMXBs evolutionary channel successfully explains the majority
of IMBPs. However, once the IMXBs evolve into LMXBs phase, the
binary orbits should widen because the material is transferred
from the less massive secondary to the more massive NS. Therefore,
this evolutionary channel cannot produce short orbital-period
($\la 3~\rm d$) IMBPs \citep{taur00}\footnote{As a possible path,
an IMXB can evolve into a compact IMBP by anomalous magnetic
braking of Ap/Bp donor star \citep{just06,shao12}.}. Nowadays,
there exist 18 known IMBPs (their orbital-periods range from 0.4
to 40 d, see also Table 2), in which 9 compact IMBPs have an
orbital-period of $\la 3$ d. The origin of compact IMBPs still
remains an unresolved puzzle.

The CO WDs in IMBPs can be formed by two evolutionary channels as
follows \citep{heuv94}: (i) A normal hydrogen-rich star burns
helium and hydrogen shells around its degenerate CO core when it
is on asymptotic giant branch phase. (ii) A He star with a mass of
$\la 2~M_{\odot}$ burns its helium shell \citep{habe86}. The
former route is an IMXB evolutionary channel, and the latter is a
NS + He star evolutionary channel. Most of NS + He star binaries
possess a compact orbit (their orbital-periods range from 0.01 to
1.0 d, see also Figure 1 in Chen et al. 2011) because of the
common-envelope evolution, and are potential progenitors of
compact IMBPs. Recently, \citet{chen11} suggested that the
progenitor of the short orbital-period IMBP PSR J1802-2124 may be
an NS + He star binary with a He star of $1.0~M_{\odot}$ and an
orbital-period of 0.5 d. Meanwhile, \citet{taur12} argued that, in
post common envelope binaries case BB RLOF of He stars can form
most binary millisecond pulsars with CO WD companions. In this
work, we attempt to systemically investigate the initial parameter
space of NS + He star binaries that could form IMBPs.

\section{Description of binary evolution}

\subsection{Stellar evolution code}
To explore the formation of double NSs binaries, \cite{dewi02} and
\cite{dewi03} have calculated the evolution of NS +
intermediate-mass He star binaries. Similar to their work, we used
an updated version of the stellar evolution code developed by
Eggleton (1971,1972,1973; see also Han et al. 1994; Pols et al.
1995) to calculate the evolutionary sequences of NS (of mass
$M_{\rm NS}$) + He star (of mass $M_{\rm He}$) binaries, until the
systems evolve into the detached binaries. The initial chemical
abundance of the He stars was taken to be Y = 0.98, Z = 0.02, and
the ratio of the mixing length to the pressure scale height and
the convective overshooting parameter were 2.0, and 0
\citep{wang09}, respectively. In addition, the stellar OPAL
opacity table originated from a version given by \cite{chen07} (e.
g. Rogers \& Iglesias 1992;  Alexander \& Ferguson 1994).

\subsection{Mass transfer and angular momentum loss}
During the evolution of binaries, orbital angular momentum loss
play a vital role. In calculation, we consider three types of
angular momentum loss from the binary system, which are described
as follows:

1. Gravitational wave radiation (GWR). For He star-NS binaries
with a compact orbit, GWR can effectively carry away orbital
angular momentum and leads to mass transfer. This angular momentum
loss rate is
\begin{equation}
\dot{J}_{\rm GR}=-\frac{32G^{7/2}}{5c^{5}}\frac{M_{\rm
NS}^{2}M_{\rm He}^{2}M^{1/2}}{a^{7/2}},
\end{equation}
where $G$ and $c$ are the gravitational constant and the speed of
light, respectively; $M=M_{\rm NS}+M_{\rm He}$ is the total mass
of the binary, and $a$ is the binary separation.

2. Mass loss. During a rapid mass transfer, there may exist mass
and angular momentum loss due to the limitation of Eddington
accretion rate, in which $\dot{M}_{\rm Edd}\simeq 3.0\times
10^{-8}~M_{\odot}{\rm yr}^{-1}$ for He-rich accretion material.
When the mass transfer rate $\mid\dot{M}_{\rm He}\mid>\dot{M}_{\rm
Edd}$, we assumed that the mass loss rate of the system
$\dot{M}=\dot{M}_{\rm He}+\dot{M}_{\rm Edd}$ proceeds according to
isotropic re-emission \citep{sobe97}, and carries away the
specific orbital angular momentum of the NS. The orbital angular
momentum loss rate by the isotropic re-emission is given by
\begin{equation}
\dot{J}_{\rm IR}=\frac{\dot{M}M_{\rm He}^{2}}{M^{2}}a^{2}\Omega,
\end{equation}
where $\Omega$ is the orbital angular velocity of the binary.

3. Magnetic braking. Even if there exist no mass exchange, the
coupling between the magnetic field and the stellar winds can also
indirectly carry away the orbital angular momentum of binaries
\citep{verb81}. In this work, the induced magnetic braking model
given by Sills et al. (2000) was adopted.

\subsection{Spin evolution of the NS}
In the stellar evolution code, the evolution of the spin and the
surface magnetic field of NSs were also considered. In summary, we
consider three spin evolution stages in Table 1. The detailed
input physics for this subsection see also section 2.3 in
\cite{liu11}.

\begin{table}
\begin{center}
\centering \caption{ Three evolutionary phases of NSs during mass
transfer.\label{tbl-1}}
%\begin{threeparttable}
\begin{tabular}{cccc}
\hline
No. & criteria & evolutionary phase & spin evolution \\
\hline
1& $r_{\rm m}<r_{\rm {c}}$ & accretion phase & spin-up\\
2& $r_{\rm m}>r_{\rm {c}}$ &  propeller phase & spin-down\\
3& $r_{\rm m}>r_{\rm {lc}}$ &  radio phase& spin-down\\

 \hline
\end{tabular}
%   \begin{tablenotes}
     \item Note: $r_{\rm m}, r_{\rm c}$, and $r_{\rm {lc}}$ represent the magnetosphere radius, the co-rotation radius, and
     the light cylinder radius of the NS, respectively.
%    \end{tablenotes}
%\end{threeparttable}
\end{center}
\end{table}

\begin{figure}
 \includegraphics[width=0.5\textwidth]{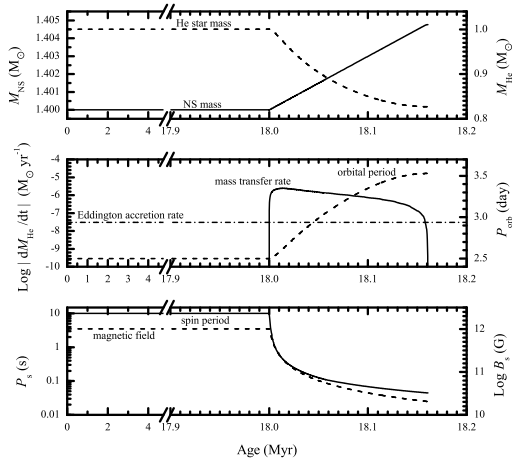}
  \caption{Evolutionary tracks of
an NS + He star binary with $M_{\rm He}^{\rm i} = 1.0 M_\odot$,
and $P_{\rm orb}^{\rm i}$ = 2.5 d. The solid and dashed curves
represent the evolution of the NS mass and the He star mass in the
top panel, the mass transfer rate and the orbital period in the
middle panel, and the spin period and the surface magnetic field
of the NS in the bottom panel, respectively. }
 \label{fig:fits}
\end{figure}

\section{Simulated results}
To explore the initial parameter space of NS + He star binaries
that can form IMBPs with various spin-periods, we have calculated
the evolution of a large number of NS + He star binaries.
Employing the population synthesis approach, \citet{chen11} found
that the initial He star masses and the initial orbital-periods of
NS + He star binaries are  $M_{\rm He}^{\rm i}\sim0.5 -
2.0~M_\odot$, and $P_{\rm orb}^{\rm i}\sim0.01 - 10 ~\rm d$,
respectively. According to their results, we take the simulated
grids to be $M_{\rm He}^{\rm i}= 0.5 - 2.0~M_{\odot}$, and $P_{\rm
orb}^{\rm i}= 0.06 - 10.0$ d. For normal radio pulsars, the
maximum spin period is 11 s \citep{man04}, and the final spin
periods of NSs is insensitive to the initial spin-periods and the
initial magnetic fields \citep{wang11}. Therefore, the initial
spin-period and the initial magnetic field of the NSs are taken to
be 10 s, and $10^{12}$ G, respectively. Furthermore, a canonical
NS mass $M_{\rm NS}^{\rm i}=1.4~M_{\odot}$ was adopted.

In Figure 1, we show the detailed evolutionary sequences of a NS +
He star binary with an initial mass $M_{\rm He}^{\rm
i}=1.0~M_{\odot}$, and an initial orbital period $P_{\rm orb}^{\rm
i}=2.5~\rm d$. After the exhaustion of central He, the He star
begins case BB mass transfer at the age of $t\simeq \rm 18.0~Myr$.
The rapid transfer of He-rich material occurs at a high rate of
$\sim 10^{-6 }~M_{\odot}\,\rm yr^{-1}$. Because of the Eddington
accretion rate, a majority of the transferred material is ejected
in the vicinity of the NS, and form the isotropic re-emission.
When $t=18.16$ Myr, the mass transfer ceases, and the binary
becomes a detached system. Hereafter, the He star evolves into a
CO WD and begins cooling. After mass transfer of  $ 0.16~\rm Myr$,
the accreted mass for the NS $\bigtriangleup M_{\rm NS}= M_{\rm
Edd}\times0.16~{\rm Myr}=0.0048~M_{\odot}$. Though the accreted
mass is $\sim3\%$ of the transferred matter, it is sufficient to
spin the NS up to $P_{\rm s}\approx 45$ ms, and causes the
magnetic field to decay to $2\times 10^{10}$ G. Owing to the less
massive He star transfers the material to the more massive NS, the
orbital period continuously increases to 3.53 d.

Figure 2 compiles the final evolutionary fate of NS + He star
binaries in the initial He star mass - orbital period ($M_{\rm
He}^{\rm i} - P_{\rm orb}^{\rm i}$) plane. It is seen that NS + He
star binaries with He stars of $0.8-1.4~M_{\odot}$ can evolve into
IMBPs. When the initial orbital period $P_{\rm orb}^{\rm i}\leq
0.1~\rm d$, most evolutionary products are IMBPs with a
millisecond period. If $P_{\rm orb}^{\rm i}> 0.1~\rm d$, nearly
all NSs are only mildly recycled, and their spin-period are in the
range of 10 -200 ms. In particular, in our calculated grids, there
exist 4 NS + He star binaries that produce slow spin pulsars with
a spin period of $200 - 1000$ ms. When the mass ratio of binaries
$q=M_{\rm He}/M_{\rm NS}>1$, the envelopes of the He stars is
dominated by convection at the onset of RLOF \citep{heuv09}.
Therefore, a runaway mass transfer is triggered, and nearly all NS
+ He star binaries with $q>1$ would experience CE evolution.
However, for H main sequence companions in LMXBs or IMXBs with
radiative (or only slightly convective) envelopes, and the mass
transfer is stable, and can avoid CE evolutionary stage even if
$q>1$ \citep[see also][]{taur99,taur00,pods02,liu11,shao12}.

In order to study how the accreted mass influences the spin
evolution of NSs, in Figure 3 we plot the evolutionary results in
the accreted mass vs. the final spin period of NSs
($\bigtriangleup M_{\rm NS} - P_{\rm s}$) diagram. It is clear
that, if NSs accrete He-rich material of $\ga~0.02~M_{\odot}$,
they will be spun up to millisecond period, and evolve into IMBPs
like PSR J1614-2230 \citep{demo10}. As shown in this figure, there
exist a practically linear logarithmic inverse correlation between
the final spin-period and the accreted mass of NSs, which
originates from our model assumption for the spin-up of NSs (see
equation 9 in Liu \& Chen 2011).

\begin{figure}
 \includegraphics[width=0.5\textwidth]{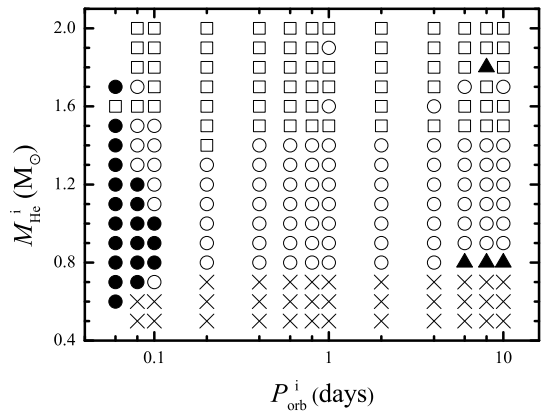}
  \caption{Distribution of the
initial He star mass $M_{\rm He}^{\rm i}$ and the initial orbital
period $P_{\rm orb}^{\rm i}$  of NS + He star binaries that can
evolve into IMBPs via the He stars evolutionary channel. The
filled circles, open circles, and filled triangles denote IMBPs
with spin-period of $1 - 10$ ms, $10 - 200$ ms, and $200 - 1000$
ms, respectively. The open squares correspond to NS + He star
binaries that experience dynamically unstable mass transfer, and
the crosses represent the binaries in which the He star does not
fill its Roche lobe in the Hubble time.}
 \label{fig:fits}
\end{figure}

In Figure 4,we present the distribution of our simulated results
in the $P_{\rm orb,f} - P_{\rm s}$ diagram. One can see that our
NS + He star evolutionary channel can form IMBPs with an orbital
period of 0.03 - 20 d and a spin-period of $1.4 - 1000$ ms. It is
difficult for our evolutionary scenario to produce IMBPs with a
relatively long orbital-period ($\ga 20$ d), which should be
derived from the selection of the distribution of the initial
orbital-periods. In Table 2, we list the observed parameters for
18 IMBPs. To compare with observations, we also show the locations
of 18 IMBPs by the filled stars in Figure 4. As shown in this
figure, our calculated results can roughly fit 5 - 6 observed
sources. Other sources have much shorter spin period than that
predicted by the model. However, for 9 short orbital-period IMBPs,
our evolution results can fit 4 sources, suggesting that the NS +
He star evolutionary channel may be an interesting evolutionary
path to short orbital-period IMBPs.

\begin{table}
\begin{center}
\centering \caption{ Observed parameters for 18
IMBPs.\label{tbl-1}}
%\begin{threeparttable}
\begin{tabular}{cccc}
\hline
Pulsars & $P_{\rm s}(\rm ms)$ &  $ P_{\rm orb}({\rm days})$ & ${\rm References}$\\
\hline
J1420$-$5625 & 40.3   & 34.1   &1 \\
J1810$-$2005 & 32.8   & 15.01  &2 \\
J1904+0412   & 71.1   & 14.93  &2 \\
J1454$-$5846 & 45.2   & 12.42  &2  \\
J1614$-$2230 & 3.15   & 8.69  & 3,4\\
J0621+1002   & 28.9   &8.319  &5,6\\
J1022+1001   & 16.5   & 7.805 &6  \\
J2145$-$0750 & 16.1   & 6.839 &7  \\
J1603$-$7202 & 14.8   & 6.309 &8   \\
J1157$-$5112 & 43.6   & 3.507 &9  \\
J1528$-$3146 & 60.8   & 3.18  &10  \\
J1439$-$5501 & 28.6   & 2.12   & 11,12  \\
J1232$-$6501 &  88.3  &  1.86 &2,13\\
J1435$-$6100 & 9.35   &  1.35 &2,13\\
B0655$+$64   &  195.7 &  1.03 &14,15\\
J1802$-$2124 & 12.6   & 0.699 &11,16\\
J1757$-$5322 & 8.87   & 0.453 &17\\
J1952+2630$^{c}$&20.7 & 0.392 &18\\
\hline
\end{tabular}
   \begin{tablenotes}
     \item References. (1)\cite{hobb04};
     (2)\cite{cami01}; (3)\cite{craw06}; (4)\cite{demo10}; (5)\cite{spla02}; (6)\cite{cami96}; (7)\cite{bail94}; (8)\cite{lori96};
     (9)\cite{edwa01a}; (10)\cite{jaco07}; (11)\cite{faul04};
     (12)\cite{lori06}; (13)\cite{man01}; (14)\cite{jone88}; (15)\cite{lori95};
     (16)\cite{ferd10}; (17)\cite{edwa01b};
     (18)\cite{knis11} .
    \end{tablenotes}
%\end{threeparttable}
\end{center}
\end{table}

\section{Discussion and summary}
In this paper, we attempt to investigate if IMBPs can have evolved
from NS + He star evolutionary channel. We have performed
numerical calculations for the evolution of NS + He star binaries
consisting of a $1.4~M_{\odot}$ NS and a 0.5 - 2.0 $M_{\odot}$ He
star companion. Our main results can be summarized as follows.

1. When the He star mass is in the range of $0.8 - 1.4~M_{\odot}$,
NS + He star binaries can evolve into IMBPs. If the initial
orbital period  $P_{\rm orb}^{\rm i}$ is too short ($< 0.1~\rm
d$), the NS will be fully recycled, and can be spun up to $<~10$
ms. In other cases, a mildly recycled pulsar with a spin-period of
10 -1000 ms would be produced.

\begin{figure}
 \includegraphics[width=0.5\textwidth]{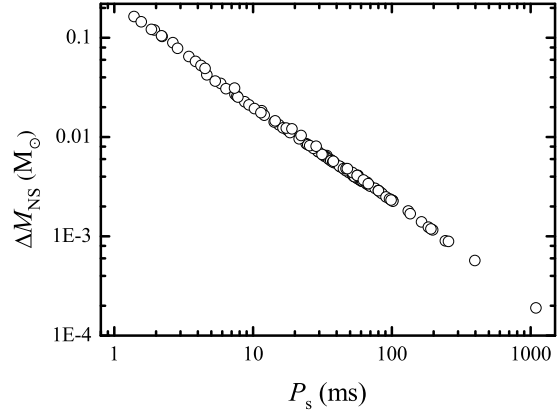}
  \caption{ Distribution of our
simulated results in the accreted mass of NSs vs. the final
spin-period ($\bigtriangleup M_{\rm NS} - P_{\rm s}$) diagram.}
 \label{fig:fits}
\end{figure}

2. During the mass exchange, the accreted mass of most NSs is in
the range of $0.001 - 0.18~M_{\odot}$, which is sufficient to spin
up the NS to a period of $1 - 200$ ms. An accreted mass of
$0.02~M_{\odot}$ is the threshold that the NSs are recycled to be
millisecond period.

\begin{figure}
 \includegraphics[width=0.5\textwidth]{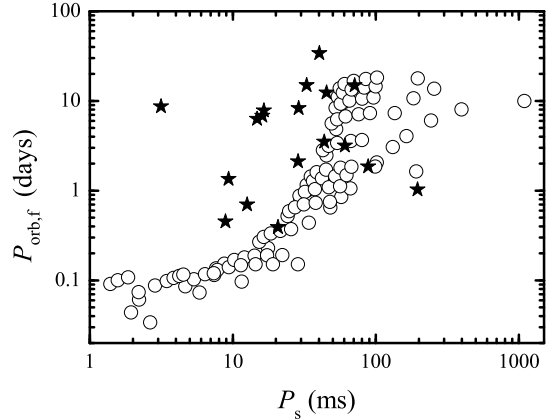}
  \caption{ Distribution of our
simulated IMBPs in the final orbital periods vs. the spin-period
of NSs ($P_{\rm orb, f} - P_{\rm s}$) diagram. The solid stars
represent the observed sources.}
 \label{fig:fits}
\end{figure}

3. The NS + He star evolutionary channel can produce IMBPs with an
orbital-period of $0.03 - 20$ d. Out of 9 known short
orbital-period IMBPs, our model can address the formation of 4
compact sources. Comparing with the IMXBs evolutionary channel,
the NS + He star binaries can form short orbital-period ($\la 3$
d) IMBPs without invoking anomalous magnetic braking of Ap/Bp
stars \citep{just06,shao12}

4. Comparing with the observations, our simulated IMBPs tend to
have a long spin-period, or a short orbital-period. There may be
three possibilities to explain the difference between observations
and our model predictions. Firstly, the NS + He star evolutionary
channel only produces a part of observed IMBPs, and the remainder
descended from IMXBs \citep{taur00,shao12}. Secondly, the
theoretical model of magnetic braking needs to be revised, or
super-Eddington accretion of NSs may occur \citep{bege02}.
Thirdly, in our calculation we take a canonical NS mass (of mass
$1.4~M_{\odot}$), while the NS masses may have a large range of
values. Actually, the initial parameter space (especially initial
orbital-period) for forming IMBPs in $P_{\rm orb}^{\rm i}-M_{\rm
He}^{\rm i}$ diagram increases with the increase of the NS mass
\citep{shao12}. Observational and theoretical studies indicate
that NSs formed by iron-core collapse supernovae and
electron-capture supernovae have obviously different initial mass
\citep[see
also][]{nomo84,timm96,hege03,woos05,schw10,zhan11,ozel12}. For
example, millisecond pulsar PSR J1614- 2230 with a mass of
$2.0~M_{\odot}$ was proposed to be originated from IMXB with a
massive NS of $1.6 - 1.7~M_{\odot}$ \citep{lin11,taur11}.

\section*{Acknowledgments}
This work was partly supported by the National Science Foundation
of China (under grant number 11173018), and Innovation Scientists
and Technicians Troop Construction Projects of Henan Province,
China.

\bsp

\label{lastpage}

\end{document}